\documentstyle{article}
\begin{document}

\begin{center}

{\Large \bf  Towards a Natural Representation of Quantum Theory: } \\
{\Large \bf  I. The Dirac Equation Revisited }

\vspace{1.5cm}

       Francesco A. Antonuccio

\vspace{3mm}

{\it Department of Theoretical Physics,}\\
{\it 1 Keble Road, Oxford OX1 3NP, United Kingdom }\\

\vspace{12mm}

{\large ABSTRACT }

\end{center}

\vspace{3mm}

An alternative (yet equivalent) formulation of the Dirac equation and
corresponding 4-spinor is investigated by considering a representation
of the Lorentz Group over a simple  {\it non-division} algebra. The
(commutative) algebra admits a natural operation of `conjugation' ,
and `unitarity' can easily be defined.
In contrast to the usual complex representation of the Lorentz group,
the one introduced here turns out to be unitary (with respect to
the algebra).

\newpage

\newcommand{\cn}{\overline}
\newcommand{\D}{${\bf D}$}
\newcommand{\beq}{\begin{equation}}
\newcommand{\eeq}{\end{equation}}
\newcommand{\ifff}{\Leftrightarrow}
\newcommand{\defn}{ \stackrel{\rm def}{=}}
\newcommand{\dw}{\Delta w}
\newcommand{\db}{{\partial}}
\newcommand{\Dn}{{\bf D}^n}
\newcommand{\Dm}{{\bf D}^m}
\newcommand{\ri}{{\mbox{i}}}
\newcommand{\jj}{{\mbox{j}}}

\section{Introduction and Motivation}

This article is motivated by the observation that the Dirac equation
can be concisely formulated without any explicit reference to
complex-valued quantities. To see this, consider the following:

\subsection{Realising the Dirac Equation}
If
\beq
     \Psi_{\bf C} = \left( \begin{array}{c}
                            u_1 + \ri v_1 \\
                            u_2 + \ri v_2 \\
                            u_3 + \ri v_3 \\
                            u_4 + \ri v_4
                           \end{array}
                    \right)
\eeq
is a Dirac 4-spinor satisfying the Dirac equation\footnote{ The {\it chiral}
representation for the $\gamma^\mu$ matrices is assumed here. See
\cite{ryder} for details.}

\beq
        ( \ri \gamma^\mu \db_{\mu} - m) \Psi_{\bf C} = 0 , \label{DiracEqn}
\eeq
then the real functions $u_i,v_i (i=1,\dots,4)$ satisfy the following
eight differential equations:

\beq
 \begin{array}{cc}
   \db_t u_1 = -\db_x u_2 - \db_y v_2 - \db_z u_1 + mv_3,   & \hspace{4mm}
   \db_t u_3 = \db_x u_4 + \db_y v_4 + \db_z u_3 + mv_1,    \\
   \db_t v_1 = -\db_x v_2 + \db_y u_2 - \db_z v_1 - mu_3,   & \hspace{4mm}
   \db_t v_3 =  \db_x v_4 - \db_y u_4 + \db_z v_3 - mu_1,   \\
   \db_t u_2 = -\db_x u_1 + \db_y v_1 + \db_z u_2 + mv_4,   & \hspace{4mm}
   \db_t u_4 = \db_x u_3 - \db_y v_3 - \db_z u_4 + mv_2 ,   \\
   \db_t v_2 = -\db_x v_1 - \db_y u_1 + \db_z v_2 - mu_4,   & \hspace{4mm}
   \db_t v_4 =  \db_x v_3 + \db_y u_3 - \db_z v_4 - mu_2.
 \end{array} \label{Dirac8}
\eeq
Curiously, there is a commutative algebra closely related to the
real numbers which admits a similarly concise formulation of the
equations listed above. Some basic definitions concerning this
algebra are presented in the next section.

\subsection{The Semi-Complex Number System}
In this article, we will be considering `numbers' of the form
\beq
                   w = t + \jj x  ,
\eeq
where $t$ and $x$ are real numbers, and $\jj$ is a commuting variable
satisfying the relation
\beq
                   \jj ^2 = 1.
\eeq
We call $t$ and $x$ the {\it real} and {\it imaginary} part
 of $w = t + \jj x$ (respectively).

\noindent
Addition, subtraction, and multiplication\footnote{One can also easily
define division; see section \ref{conjugation}.}
 are defined in the obvious
way :
\begin{eqnarray}
 (t_1 + \jj x_1) \pm (t_2 + \jj x_2) & = & (t_1 \pm t_2) +
                                               \jj (x_1 \pm x_2),\\
 (t_1 + \jj x_1) \cdot (t_2 + \jj x_2) & = & (t_1 t_2 + x_1 x_2)+
                                            \jj (t_1 x_2 + x_1 t_2).
\end{eqnarray}
The zero element is $0 = 0 + \jj 0$, and two numbers $t_1+\jj x_1$,
$t_2+\jj x_2$ are equal if and only if $t_1 = t_2$ and $x_1 = x_2$.

The set of all such numbers will be denoted by the symbol \D, and we
shall refer to this set as the {\it semi-complex number system}\footnote{
The more familiar term  {\it hyperbolic quasi-real numbers}
is avoided for the sake of brevity.} For a more detailed
account of this number system, the reader is referred to \cite{czech},
\cite{fool}.

\subsection{Dirac's Equation Revisited}

Define four (real) $4 \times 4$ matrices, $\xi^\mu$, ($\mu = 0,\ldots ,3$)
by writing

\beq
 \begin{array}{cccc}
   \xi^0 = \left(
            \begin{array}{cc}
              0 & \mbox{{\bf 1}} \\
              -\mbox{{\bf 1}} & 0
            \end{array}
           \right), & \hspace{3mm}

 \xi^1 = \left(
            \begin{array}{cc}
             \mbox{{\bf 1}} & 0 \\
              0 & -\mbox{{\bf 1}}
            \end{array}
           \right), & \hspace{3mm}

 \xi^2 = \left(
            \begin{array}{cc}
              0 & \tau_1 \\
              \tau_1 & 0
            \end{array}
           \right), & \hspace{3mm}

 \xi^3 = \left(
            \begin{array}{cc}
              0 & \tau_3 \\
              \tau_3 & 0
            \end{array}
           \right),
 \end{array}
\eeq
where $\tau_1 =
 \left( \begin{array}{cc} 0 & 1 \\ 1 & 0 \end{array} \right)$,
 $\tau_3 =
\left( \begin{array}{cc} 1 & 0 \\ 0 & -1 \end{array} \right)$,
 and $\mbox{{\bf 1}}$ is the $2 \times 2$ identity matrix. We choose
this notation  for later convenience. The anti-commutation relations
for the $\xi^\mu$ matrices take the form
\beq
           \{ \xi^\mu, \xi^\nu \} = - 2g^{\mu \nu}, \hspace{5mm}
                       \mu,\nu = 0,\dots,3.
\eeq

We now introduce the semi-complex analogue of the Dirac 4-spinor: Let
\beq
   \Psi_{{\bf D}} = \left( \begin{array}{c}
                            a_1 + \jj b_1 \\
                            a_2 + \jj b_2 \\
                            a_3 + \jj b_3 \\
                            a_4 + \jj b_4
                           \end{array}     \label{SC4spinor}
                    \right),
\eeq
where $a_i,b_i$ ($i=1,\ldots ,4$) are real, and suppose
$\Psi_{{\bf D}}$ satisfies the equation
\beq
           (\jj \xi^\mu \db_\mu - m)\Psi_{{\bf D}} = 0.  \label{SCEqn}
\eeq
Separating the real and imaginary parts of this last equation yields
a set of eight real partial differential equations connecting the real
functions $a_i,b_i$. Interestingly, if we make the substitutions
\beq
 \begin{array}{ll}
     a_1 \rightarrow u_2 & \hspace{5mm} b_1 \rightarrow v_3 \\
     a_2 \rightarrow v_1 & \hspace{5mm} b_2 \rightarrow -u_4 \\
     a_3 \rightarrow u_1 & \hspace{5mm} b_3 \rightarrow -v_4 \\
     a_4 \rightarrow v_2 & \hspace{5mm} b_4 \rightarrow u_3 ,
  \end{array}
\eeq
then these eight differential equations become identical to the set
of equations listed in (\ref{Dirac8}). In other words, if we make the
identification
\beq
 \begin{array}{ccc}
   \Psi_{{\bf C}} = \left( \begin{array}{c}
                            u_1 + \ri v_1 \\
                            u_2 + \ri v_2 \\
                            u_3 + \ri v_3 \\
                            u_4 + \ri v_4
                           \end{array}
                    \right) & \leftrightarrow &
   \Psi_{{\bf D}} = \left( \begin{array}{c}
                            u_2 + \jj v_3 \\
                            v_1 - \jj u_4 \\
                            u_1 - \jj v_4 \\
                            v_2 + \jj u_3
                           \end{array}
                    \right),
    \end{array}
\eeq
then  equation (\ref{DiracEqn}) (Dirac's equation), and
 equation (\ref{SCEqn}), are
entirely equivalent. The suggestion here is that the Dirac equation
is not a fundamentally `complex' equation.

Fortunately, this observation turns out to be more than just a curious
accident, and a better understanding can be obtained by investigating
a particular representation of the Lorentz Group.
This topic will be taken up  next.

\newpage
\section{The Lorentz Algebra}
\subsection{A Complex Representation}

Under Lorentz transformations, the Dirac 4-spinor $\Psi_{{\bf C}}$
transforms as follows:
\beq
 \Psi_{{\bf C}} \rightarrow
         \left( \begin{array}{cc}
                   e^{\frac{i}{2} \sigma \cdot (\theta - i \phi)} & 0 \\
                   0 & e^{\frac{i}{2} \sigma \cdot (\theta + i \phi)}
                 \end{array}
         \right) \Psi_{{\bf C}} . \label{LorentzC}
\eeq
The six real parameters $\theta = (\theta_1, \theta_2,\theta_3 )$ and
$\phi = ( \phi_1, \phi_2, \phi_3)$ correspond to  three generators
for spatial rotations, and three for Lorentz boosts respectively. The
matrices $\sigma = ( \sigma_x, \sigma_y, \sigma_z )$ are the well
known Pauli matrices.

\medskip
\noindent
Let us introduce six matrices $E_i, F_i$ $(i = 1,2,3)$ by writing
\beq
 \begin{array}{lll}
    E_1 = \frac{1}{2} \left( \begin{array}{cc}
                               \sigma_x & 0 \\
                                 0 & -\sigma_x
                             \end{array}
                      \right) & \hspace{4mm}
    E_2 = -\frac{\ri}{2} \left( \begin{array}{cc}
                               \sigma_y & 0 \\
                                 0 & \sigma_y
                                 \end{array}
                      \right) & \hspace{4mm}

    E_3 = \frac{1}{2} \left( \begin{array}{cc}
                               \sigma_z & 0 \\
                                 0 & -\sigma_z
                                \end{array}
                      \right) \\  & & \\

    F_1 = \frac{\ri}{2} \left( \begin{array}{cc}
                               \sigma_x & 0 \\
                                 0 & \sigma_x
                                \end{array}
                      \right) & \hspace{4mm}

    F_2 = \frac{1}{2} \left( \begin{array}{cc}
                               \sigma_y & 0 \\
                                 0 & -\sigma_y
                               \end{array}
                      \right) & \hspace{4mm}

    F_3 = \frac{\ri}{2} \left( \begin{array}{cc}
                               \sigma_z & 0 \\
                                 0 & \sigma_z
                                \end{array}
                      \right)
 \end{array} \label{EFC}
\eeq
Then transformation (\ref{LorentzC}) may be written as follows:
\beq
 \Psi_{{\bf C}} \rightarrow
     \exp(\phi_1 E_1 - \theta_2 E_2 + \phi_3 E_3 + \theta_1 F_1 +
                         \phi_2 F_2 + \theta_3 F_3 ) \Psi_{{\bf C}}.
\eeq
The algebra of commutation relations for the  matrices $E_i,F_i$
is given below:
\beq
 \begin{array}{llll}
  [ E_1, E_2 ] = E_3 & \hspace{8mm}  [F_1,F_2] = - E_3  &  \hspace{8mm}
  [E_1,F_2] = F_3  & \hspace{8mm}   [F_1,E_2]=F_3 \\ \mbox{ }
 [ E_2, E_3 ] = E_1 &  \hspace{8mm} [F_2,F_3] = - E_1 &  \hspace{8mm}
  [E_2,F_3] = F_1  & \hspace{8mm}   [F_2,E_3]=F_1 \\ \mbox{ }
 [ E_3, E_1 ] = -E_2 & \hspace{8mm}  [F_3,F_1] = E_2  &  \hspace{8mm}
  [E_3,F_1] = -F_2  &  \hspace{8mm}  [F_3,E_1]= - F_2
 \end{array} \label{LorentzAlg}
\eeq
All other commutators vanish. Abstractly, these commutation relations
define the Lie algebra of the Lorentz Group $O(1,3)$, and the complex
matrices $E_i,F_i$ defined by (\ref{EFC}) correspond to a {\it complex}
representation of this algebra.

\subsection{A Semi-Complex Representation}

It turns out that there exists a {\it semi-complex} representation
of the Lorentz algebra (\ref{LorentzAlg}). In order to obtain an
explicit presentation, we proceed as follows:

\medskip
\noindent
Define three $2 \times 2$ matrices $\tau = (\tau_1,\tau_2,\tau_3)$ by
setting
\beq
 \begin{array}{lll}
   \tau_1 = \left(
              \begin{array}{cc}
                0 & 1 \\
                1 & 0
               \end{array}
             \right), & \hspace{8mm}
 \tau_2 = \left(
              \begin{array}{cc}
                0 & -\jj \\
                \jj & 0
               \end{array}
             \right), & \hspace{8mm}
 \tau_3 = \left(
              \begin{array}{cc}
                1 & 0 \\
                0 & -1
               \end{array}
             \right).
 \end{array}        \label{tau}
\eeq
The reader may like to verify the following commutation relations for
the matrices $\tau_i$ :
\beq
 \begin{array}{lll}
   [\tau_1 ,  \tau_2 ] = 2 \jj \tau_3, &  \hspace{7mm}
   [\tau_2 ,  \tau_3 ] = 2 \jj \tau_1, &  \hspace{7mm}
   [\tau_3 ,  \tau_1 ] = -2 \jj \tau_2 .
  \end{array}
\eeq
Now redefine the matrices $E_i,F_i$ $(i = 1,2,3)$ by setting
\beq
  E_i = \frac{\jj}{2} \left(
                        \begin{array}{cc}
                          \tau_i & 0 \\
                           0  & \tau_i
                         \end{array}
                       \right), \hspace{9mm}
  F_i = \frac{1}{2} \left(
                        \begin{array}{cc}
                          0 & \tau_i  \\
                          -\tau_i & 0
                         \end{array}
                       \right), \hspace{7mm} i=1,2,3.  \label{SCrep}
\eeq
A straightforward calculation shows that these matrices do indeed satisfy the
Lorentz algebra defined by the commutation relations
(\ref{LorentzAlg}). Consequently, if $ \Psi_{{\bf D}}$ is a
semi-complex 4-spinor (as in expression (\ref{SC4spinor})), then under
Lorentz transformations, $ \Psi_{{\bf D}}$ transforms as follows:
\beq
 \Psi_{{\bf D}} \rightarrow
     \exp(\phi_1 E_1 - \theta_2 E_2 + \phi_3 E_3 + \theta_1 F_1 +
                         \phi_2 F_2 + \theta_3 F_3 ) \Psi_{{\bf D}},
              \label{SCtrans}
\eeq
where this time, the matrices $E_i,F_i$ are semi-complex.

We shall discover in the next section that the semi-complex
representation  has the distinct advantage of
being {\it unitary}. This means that the exponential in
transformation (\ref{SCtrans}) is a (semi-complex) unitary matrix,
which we shall define next. In fact,
since the semi-complex algebra
admits a very natural operation of `conjugation', the
definition of unitarity presented in the next section should look
very familiar, and very natural.

\section{Unitarity}
We shall endeavour to present an elementary (i.e. brief)
introduction to  the semi-complex unitary
groups. We begin by defining `conjugation' for semi-complex numbers.

\subsection{Conjugation}
\label{conjugation}
 Given
any semi-complex number $w=t+jx$, we define the {\em conjugate} of $w$,
written $\cn{w}$,  to be
\[ \cn{w} = t-jx . \]
Two simple consequences can be immediatetely deduced; for
any $w_1,w_2 \in$ \D, we have
\begin{eqnarray}
 \overline{w_1+w_2} & = & \overline{w_1}+\overline{w_2} \hspace{4mm}
                           \mbox{and} \\
 \overline{w_1 \cdot w_2} & = & \overline{w_1} \cdot \overline{w_2}.
\end{eqnarray}
We also have the  identity
\beq
\cn{w} \cdot w = t^2 - x^2.
\eeq
Hence $\cn{w} \cdot w$ is {\em real} for any semi-complex
number $w$, although unlike the complex case, it may
take on {\em negative } values. In order to strengthen the analogy
between the semi-complex and complex numbers, we often write
\[ |w|^2 = \cn{w} \cdot w  \]
where $|w|^2$ is referred to as the `modulus squared'
of $w$. A nice  consequence of these definitions
can now be stated:
For any semi-complex numbers $w_1$, $w_2 \in$ {\bf D},
\[ |w_1 \cdot w_2|^2 = |w_1|^2 \cdot |w_2|^2 . \] \label{propmult}

Now observe that if $|w|^2$ does not vanish, the quantity
\beq
  w^{-1} = \frac{\overline{w}}{|w|^{2}}
\eeq
is a well defined inverse for $w$. So $w$ fails to have an
inverse if (and only if)  $|w|^2 = t^2-x^2 = 0$.

\subsection{The Semi-Complex Unitary Groups}

{\bf Hermiticity}: Suppose $H$ is a matrix of arbitrary dimensions
with semi-complex entries. The {\it adjoint} of $H$, written
$H^{\dagger}$, is obtained by transposing $H$, and then conjugating
each of the entries. We say $H$ is {\it Hermitian} if $H^{\dagger} = H$,
and {\it anti-Hermitian} if $H^{\dagger} = -H$. For example, the
$\tau_i$ matrices defined in (\ref{tau}) are Hermitian.

\medskip
\noindent
{\bf Unitarity}: The semi-complex {\it unitary group} U($n,{\bf D}$) is
the set of all $n \times n$ semi-complex matrices $U$ satisfying the
identity
\beq
                   U^{\dagger} U = 1.
\eeq
The {\it special unitary group} SU($n,{\bf D}$)  is defined to be the
set of all elements $U\in$U($n,{\bf D}$) with unit determinant:
\beq
                   \det U = 1.
\eeq

\medskip
\noindent
{\bf Unitarity and Hermiticity}: If $H$ is an $n \times n$ Hermitian
matrix (over \D), then $e^{jH}$ is an element of the unitary group
U($n,{\bf D}$). Equivalently, $e^{H}$ is unitary if $H$ is
anti-Hermitian. If the trace of $H$ vanishes, then $e^{jH}$
(or $e^{H}$ in the anti-Hermitian case) is contained in
the special unitary group  SU($n,{\bf D}$).

\medskip
\noindent
{\it Remark}: According to these definitions, the exponential
appearing in the Lorentz transformation (\ref{SCtrans}) is
an element of the special unitary group  SU($4,{\bf D}$), since the
traceless generators $E_i,F_i$ defined in (\ref{SCrep}) are anti-Hermitian.
So the Lorentz group is just a six-dimensional (Lie) subgroup
of the fifteen dimensional Lie group SU($4,{\bf D}$).
%
\subsection{An Isomorphism: The Spin Group}
We claim that the semi-complex group  SU($2,{\bf D}$) is just
the familiar {\it complex} group SU($1,1$). In fact, if we restrict
our attention to real numbers $a_1,a_2,b_1,b_2$ satisfying the
constraint $a_1^2 + a_2^2 -b_1^2 - b_2^2 = 1$, then the identification
\beq
 \left( \begin{array}{cc}
          a_1 + \ri a_2 & b_1 + \ri b_2 \\
          b_1 - \ri b_2 & a_1 - \ri a_2
        \end{array}
 \right) \hspace{4mm}
  \leftrightarrow
         \hspace{4mm}
 \left( \begin{array}{cc}
          a_1 + \jj b_1 & -a_2 + \jj b_2 \\
          a_2 + \jj b_2 & a_1 - \jj b_1
        \end{array}
 \right)   \label{iso2}
\eeq
establishes a (group) isomorphism between  SU($1,1$) and  SU($2,{\bf D}$),
as claimed. The terminology `spin group' for SU($2,{\bf D}$)
anticipates a forthcoming article investigating the intimate
relationship between this group, spin, and Lorentz invariance in
$2+1$ space-time.

The identification (\ref{iso2}) above suggests that the conformal
group SU($2,2$) might be just the group SU($4,{\bf D}$); at any rate,
they are both fifteen dimensional. Initial investigations suggest
that the differences  between these groups might be very subtle.
A definitive proof demonstrating the (non)existence of an
isomorphism --- at least
between the associated Lie algebras --- would be an important
initial step before investigating the physics of SU($4,{\bf D}$).

\newpage
\section{The Dirac Equation Revisited II}
\subsection{Quantisation: A New Perspective}

The  Dirac equation (\ref{SCEqn})
with zero mass ($m=0$) may be written in the form
\beq
       \jj \frac{\db \Psi_{{\bf D}}}{\db t} = H  \Psi_{{\bf D}},
                              \label{shr}
\eeq
where the operator $H$ turns out to be
Hermitian\footnote{i.e.  $< H \Psi_{{\bf D}}, \Phi_{{\bf D}}> =
 <  \Psi_{{\bf D}}, H \Phi_{{\bf D}}>$.}  with respect to the
semi-complex inner product
\beq
   <  \Psi_{{\bf D}}, \Phi_{{\bf D}}> :=
            \int \Psi_{{\bf D}}^{\dagger} \Phi_{{\bf D}} d^3 x .
\eeq
So the scalar quantity $|\Psi_{{\bf D}}|^2 := \Psi_{{\bf D}}^{\dagger}
\Psi_{{\bf D}}$ may be viewed as some kind of Lorentz invariant
density function\footnote{Non-positive definiteness suggests it might
be charge density.}.

\medskip
\noindent
So how do we quantise field equations in general such as the massless
Dirac equation given by expression (\ref{shr})? Associated
with any equation of this form is a {\it propagator} $K$,
which, in the path integral formalism, necessitates the
evaluation of a path integral
\beq
        K = \int e^{jS} {\cal D} \psi^{\dagger} {\cal D} \psi .
                                  \label{pathint}
\eeq
The functional $S=S[\psi^{\dagger},\psi]$ is an appropriately
chosen action,
which we may assume to be real-valued\footnote{Say, a real Grassmann
variable, for the fermionic case}. Note that the integrand of
this path integral is a semi-complex phase $e^{jS}$, contrasting
the usual prescription of integrating over a complex phase $e^{iS}$.

Now any semi-complex number may be uniquely decomposed into a
{\it real} linear combination of the following two projections:
\beq
       p_{+} = \frac{1}{2}(1 + \jj ) \hspace{6mm}
                      p_{-} = \frac{1}{2}(1 - \jj )
\eeq
($p_{+}$ and $p_{-}$ are projections since $p_{+} + p_{-} = 1$,
$p_{+}^2 = p_{+}$, $p_{-}^2=p_{-}$, and $p_{+}p_{-}=p_{-}p_{+}=0$).

\medskip
\noindent
In particular, $e^{jS} = \cosh{S} + \jj \sinh{S}$ admits
the following decomposition:
\beq
        e^{jS} = e^{S} \cdot p_{+} + e^{-S} \cdot p_{-}.
\eeq
So the path integral (\ref{pathint}) takes the form $K= K_{+} \cdot p_{+}
+ K_{-} \cdot p_{-}$, where the projection coefficients $K_{+},K_{-}$
are given by
\begin{eqnarray}
   K_{+} & = & \int e^S {\cal D} \psi^{\dagger} {\cal D} \psi \label{p1} \\
   K_{-} & = & \int e^{-S} {\cal D} \psi^{\dagger} {\cal D} \psi \label{p2}
\end{eqnarray}
If $S$ were positive-definite (e.g. if we replace $S$ with $|S|$),
one would expect (\ref{p1}) to be
divergent, although  (\ref{p2}) might be well defined. So
the original integral (\ref{pathint}) may be undefined as a whole,
but it may nevertheless possess a well defined  {\it projection}.
Mathematically, this is an elegant way of separating unwanted
infinities.

Moreover, the semi-complex formalism appears to have removed
the troublesome complex phase in the path integral
without resorting to an analytic
continuation to Euclidean space-time. The integrity of Minkowski
space is thus preserved.

\subsection{Gauge Invariance}

Finally, we indulge in some fanciful speculation:
First, observe that transformations of the type
\beq
 \Psi_{{\bf D}} \rightarrow e^{j\theta} \cdot \Psi_{{\bf D}}
\hspace{5mm}  (\theta \in {\bf R}) \label{glob}
\eeq
leave  $|\Psi_{{\bf D}}|^2$ invariant. So the Lagrangian
\beq
 {\cal L} = {\Psi_{{\bf D}}}^{\dagger} (\jj \xi^\mu \db_\mu - m)\Psi_{{\bf D}},
\eeq
which gives rise to the Dirac equation (\ref{SCEqn}) under variation,
is also invariant under transformations of this type.
 Gauging this global
transformation (i.e. allowing $\theta$ to vary at different space-time
points)  augments the Lagrangian into
the following expression:
\beq
   {\cal L} =
  {\Psi_{{\bf D}}}^{\dagger} (\jj \xi^\mu D_\mu - m)\Psi_{{\bf D}}
       -\frac{1}{4}F_{\mu \nu}F^{\mu \nu}. \label{grav}
\eeq
Here, $D_{\mu} = \db_\mu - \jj G_\mu$ is the covariant derivative
operator,
$G_\mu$ is the associated gauge field, and
 $F_{\mu \nu} = \db_\mu G_\nu - \db_\nu G_\mu$ is the (gauge
invariant)
field tensor. The Lagrangian (\ref{grav}) is now invariant under the
gauge transformations
\begin{eqnarray}
   \Psi_{{\bf D}} & \rightarrow & e^{j\theta(x)} \cdot  \Psi_{{\bf D}} \\
    G_\mu    & \rightarrow &  G_\mu  + \db_\mu \theta .
\end{eqnarray}
The transformation (\ref{glob}) is not a Lorentz
transformation, so what can it be? The similarities between
the electromagnetic field and the gauge field $G_\mu$ are
rather striking. For example, in the source free case, the
field tensors are exactly equivalent.

In the physical world, there are two forces which, classically,
appear very similar; namely, the Coulomb attraction between two stationary
charges, and the gravitational force between two masses,
each of which are described by an
inverse square law. It is tempting, then, to view the U($1,{\bf D}$)
gauge theory just presented as a simple gauge theory of gravity.
We leave such speculations to the interested reader!

\section{Concluding Remarks}

Our observations suggest that it is unnecessary (and potentially
restrictive) to view complex-valued quantities as fundamental
to a relativistic  quantum theory.

Quantum physics depends heavily on the concept of {\it Hermitian
operators}, since it is precisely these operators which guarantee
a real (i.e. measurable) spectrum of eigenvalues. Consequently,
if we are seeking a quantum theory which is consistent with
relativity---an essential requirement for quantum gravity---then
it is appropriate that a unitary-like representation of the Lorentz
group be considered. Such a representation {\it can} be
obtained, but it requires the somewhat unorthodox step of
embracing an unfamiliar number algebra.

Hopefully, the effort of venturing beyond the familiar complex
number system will be more than compensated by a new and
fruitful perspective on the old quantum world.

\medskip
{\bf Acknowledgements}

I am indebted to many members of the Theoretical Physics and
 Mathematics Department (Oxford) for stimulating discussions on
 mathematical physics.
This work was financially supported by the Commonwealth
Scholarship and Fellowship Plan (The British Council, U.K.),
and is dedicated to I.W.



\end{document}